Extension of Relativistic-Microwave Theory of Ball Lightning Including Long-term Losses And Stability

Karl D. Stephan, Ingram School of Engineering, Texas State University, San Marcos, Texas USA

Abstract: After centuries, the long-standing problem of the nature of ball lightning may be closer to a solution. The relativistic-microwave theory of ball lightning recently proposed by Wu accounts for many of the leading characteristics of ball lightning, which most previous theories have failed to do. It involves the impact of a lightning-caused relativistic electron bunch to soil, producing an EM pulse that forms a plasma bubble. While the theory presents a plausible account of ball-lightning formation, storing electromagnetic energy long enough to account for the observed lifetime of such objects was not demonstrated. Here we show how such a structure can develop the high $Q$ factor ($\sim 10^{10}$) needed for the observed lifetimes of ~seconds for ball lightning, and show that the structure is radially stable, given certain assumptions.

Relativistic-Microwave Ball-Lightning Theory

Ball lightning is an as-yet-unexplained phenomenon of atmospheric physics [1,2] whose scientific investigation dates back to the nineteenth century. In appearance, a typical ball-lightning object is a glowing spheroid averaging about 25 cm in diameter which appears during a thunderstorm, and moves horizontally more frequently than vertically, having a lifetime on the order of seconds [3]. Although most data on the phenomenon comes from eyewitness accounts, recently Cen et al. [4] obtained videography and visible-wavelength spectra of an object that was probably ball lightning. The spectra indicated the presence of elements found in soil (Si, Ca, Fe) as well as excited air molecules and atoms.

Many theories [1] have been proposed to explain ball lightning, but most of them fail to account for more than one or two salient characteristics, and none has provided experimentalists with enough information to reproduce the phenomenon in the laboratory. Wu [5] recently proposed a theory which can potentially account for most of the principal observed characteristics of ball lightning. The main features of his theory are as follows: (1) A cloud-to-ground negative lightning leader produces a bunch of at least $\sim 10^{14}$ electrons in a compact cm-size cluster moving at a relativistic speed (>1 MeV). (2) When the bunch encounters a solid object (e. g. soil) the electromagnetic wave accompanying it is reflected and produces a pulse of intense electric fields on the order of $10^8$ V m$^{-1}$ concentrated in the microwave frequency range (~1 GHz). (3) When this pulse encounters a plasma of sufficient density, the ponderomotive force associated with such high fields clears out most electrons from the central region of the pulse and forms a spheroidal plasma bubble, trapping the field inside a self-generated microwave resonator. Wu presented PIC-code-generated simulations based on his theory which produced such structures, and cited experiments [6] in which a high-power laser produced transient micron-scale bubble-like objects in a jet of helium gas. However, the longest duration of Wu's simulation presented in [5] was only 20 ns. While several cycles of the microwave-cavity oscillation executed during that time, the evidence presented was not conclusive that the object modeled could sustain itself in existence over the 1-10 sec lifetime that typical ball-lightning observers report.

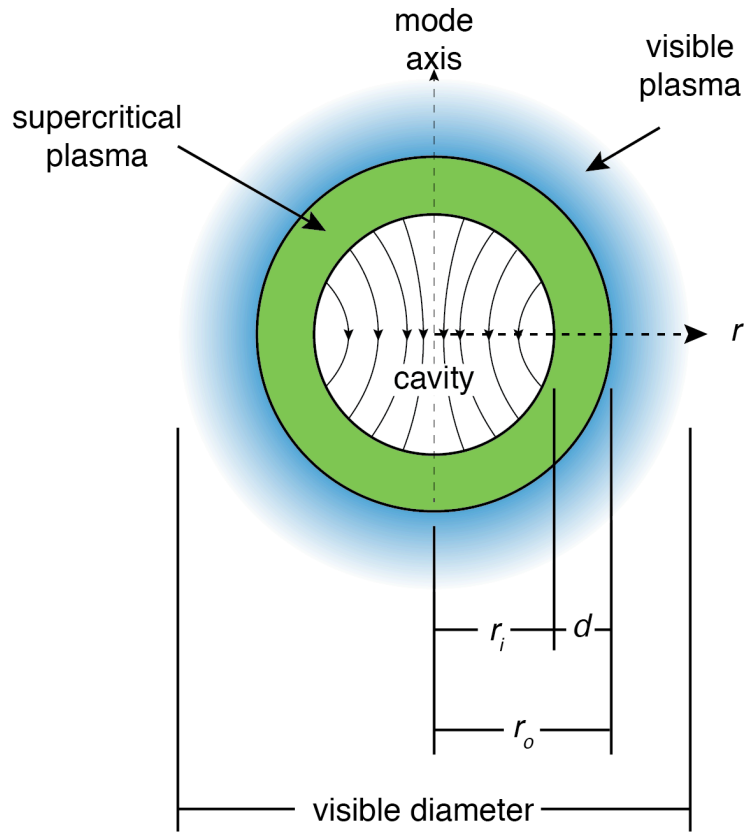

Fig. 1. Ball-lightning structure proposed and modeled by Wu, showing spherical cavity of radius $r_i$, supercritically-dense plasma region of thickness $d$, and an outer luminous layer.

Fig. 1 shows the structure of a ball-lightning object according to Wu's theory, at a time immediately after its formation. Inside the resonant cavity of radius $r_i$, there are intense electromagnetic fields corresponding to the excited modes of the cavity. (The electric field lines of the lowest-order ($TM_{101}$) mode are shown, but other modes may be present as well.) Because the ponderomotive force is more effective for electrons than for ions, initially there is a large net positive charge density inside the cavity. The cavity boundary occurs at $r_i$ where the electron number density $n_e$ exceeds the critical density $n_c$ needed for microwaves to be reflected from the plasma. Because the plasma's density between $r_i$ and $r_o$ exceeds the critical density $n_c$, we term this the supercritical plasma layer. Wu's model shows that this region can have a thickness of several cm, and is negatively charged, preserving neutrality of charge for the overall structure. As shown below, the microwave fields attenuate exponentially in the supercritical plasma shell over its thickness $d$ to a level at $r_o$ that accounts for the observed relatively weak visible-light radiation emitted by ball lightning.

Extension of Theory To Include Losses Explicitly

In this paper we assume that Wu's theory is basically correct, and combine it with observed characteristics of ball lightning to draw conclusions about the ability of such a structure to store many kJ of energy in the form of an intense oscillatory microwave electromagnetic field. We find that in terms of a cavity $Q$-factor, a simple calculation shows that $Q$ on the order of $10^{10}$ is required. We show what this requirement implies for the electron number density in the supercritical plasma shell, and demonstrate how such a plasma can act as a nearly perfect conductor by virtue of a large negative permittivity.

In 1955 Kapitza ([7], cited in [1], p. 181) was among the first to propose a detailed resonant-electromagnetic-cavity model for ball lightning. All such models face the difficulty that storing kJ levels of electrical energy in a gas or plasma on the order of a meter or less in diameter requires extremely high electric and magnetic fields, on the order of $10^8$ V m$^{-1}$ and $10^5$ A m$^{-1}$, respectively. Until now, it was not clear how these fields could be either produced or maintained without causing immediate breakdown of the air and consequent rapid dissipation of energy. However, Wu's simulations indicate that the plasma surrounding the high-field region can both reflect the fields with sufficiently low losses and shield the surrounding air enough to achieve the energy confinement required. The following elaboration of the model will show in detail how this can come about.

The fundamental definition of the quality factor $Q$ of a resonant-cavity mode is

$$Q \equiv \frac{2\pi(\text{energy stored in mode})}{\text{energy lost per cycle}} \qquad (1)$$

For a spherical cavity of radius $a$ supporting the lowest-order $TM_{101}$ mode, it can be shown [8] that the free-space resonant wavelength in the cavity medium is $\lambda_0 \sim 2.29a$. Furthermore, if the walls have a surface resistivity $R_S$ (ohms per square), the $Q$ of this mode of the spherical cavity is [8]

$$Q \approx 1.31 \frac{\eta_0}{R_S}, \qquad (2)$$

where $\eta_0$ is the impedance of free space (376.7 Ω). An estimate of the $Q$ required to account for the lifetime of a typical ball-lightning object can thus be used to estimate the effective surface resistivity required of the plasma surrounding the cavity.

A resonator oscillating at a frequency $f_0 = \omega_0/2\pi$ with a quality factor $Q$ whose oscillation amplitude at $t = 0$ is $A_0$ will show an amplitude dependence $A(t) = A_0\exp(-\alpha t)$, where

$$\alpha = \frac{\omega_0}{2Q}. \tag{4}$$

As the energy stored in the microwave fields decreases, the ponderomotive force maintaining the high electron concentration in the supercritical plasma layer also decreases, and at some point this concentration will fall below the critical value $n_c$ needed to maintain the integrity of the cavity. Wu has shown [5] that a critical electron number density $n_c = 1.2 \times 10^{10}$ cm$^{-3}$ or greater in the plasma surrounding the structure is necessary for the cavity to maintain itself at a resonant frequency of 1 GHz. This critical density is correlated with a peak oscillatory electric field within the cavity of approximately $10^8$ V m$^{-1}$. As an initial field much higher than this is not likely, we will assume for a typical case that the initial electric field intensity is three times higher, or $3 \times 10^8$ V m$^{-1}$. If the field amplitude follows the exponential time dependence $A_0\exp(-\alpha t)$, we can express the object's lifetime $\tau_{LT}$ as the time when the field will have decayed by a factor of 3 to the minimum level needed to support the cavity structure:

$$A(\tau_{LT}) = \frac{A_0}{3} = A_0 e^{-\alpha \tau_{LT}} \tag{5}$$

Using Equation (4), we can find the minimum $Q$ required, $Q_{MIN}$, in terms of the object's lifetime $\tau_{LT}$ and the mode's resonant frequency $f_0$:

$$Q_{MIN} = 2.8 f_0 \tau_{LT} \tag{6}$$

If $f_0 = 1$ GHz and object lifetime $\tau_{LT} = 5$ s (a typical value for many observations), $Q_{MIN} = 1.4 \times 10^{10}$. Although the decay factor of 3 assumed above is arbitrary, a decay factor higher than 10 leads to implausibly high values for initial field intensity, and will not reduce the required $Q$ much below $10^9$ in any case.

Microwave cavities with $Q$ values exceeding $10^{10}$ are routinely used in particle accelerators, but they are carefully fabricated superconducting cavities operating at cryogenic temperatures. A 25-cm-diameter spherical cavity made of copper sustains the $TM_{101}$ mode at about 1 GHz. At that frequency, the room-temperature surface resistivity of copper is 8.25 m$\Omega$, resulting in a $Q$ of only about 60,000. Once this mode is excited, the field amplitude in such an isolated copper cavity decreases by a factor of 3 in only 21 $\mu$s, far shorter than the typical lifetime exhibited by ball lightning. If the relativistic-microwave theory is correct, however, the plasma surrounding the electromagnetic field in a ball-lightning structure must isolate it from its surroundings with vanishingly low losses, much lower than that of copper.

As is well known, electromagnetic waves will not propagate in a plasma below the plasma frequency

$$\omega_p = \sqrt{\frac{n_e q^2}{\varepsilon_0 m_e}} \tag{7}$$

in which $q$ = electron charge, $m_e$ = electron mass, and $\varepsilon_0$ = permittivity of free space [9]. For a frequency $\omega$ lower than this, plasma shows negative polarization with a relative permittivity

$$\varepsilon_r = 1 - \frac{\omega_p^2}{\omega(\omega - i\nu_{en})}, \tag{8}$$

where $\nu_{en}$ is the electron-neutral collision frequency, representing energy losses. A collisionless plasma ($\nu_{en} = 0$) excited well below the plasma frequency is in principle a perfect reflector of electromagnetic energy, because the permittivity is negative, resulting in total reflection from the plasma surface.

Near-total reflection can occur only if the plasma thickness ($d$ in Fig. 1) is such that the field magnitude at the outside of the cavity wall (plasma) at $r_o$ is small enough to transmit negligible power compared to the reflected power incident on the inside walls of the cavity. Observations can put an approximate upper bound on this estimated external power level. Eyewitnesses providing an estimate of the brightness of ball-lightning objects sometimes compare it to an incandescent light bulb with a power rating between 20 W and 200 W [10]. Assuming an energy-conversion efficiency from total emitted power to visible light of 10%, we estimate that the total power (not just visible light) escaping from a typical ball-lightning object during its lifetime is probably in the range of 200 W to 2 kW, and possibly less. Additional eyewitness observations and analyses [11] indicate that many typical ball-lightning objects have an energy density (total energy/visible-object volume) between 1 and 10 MJ m$^{-3}$, which means a 25-cm-diameter object would store 8-80 kJ. This is more than enough energy to account for the observed light output, but only if the power density at the outer boundary of the object during the object's lifetime is limited to a small fraction of the effective power density inside the proposed resonant-cavity structure.

Both considerations of cavity Q and limited fields at the exterior of a ball-lightning object lead to the conclusion that between the assumed intense microwave fields inside the object and the ambient air outside, there must be a layer (presumably plasma) which can (a) reflect microwaves by means of an effective surface resistance (ohms per square) on the order of 35 n$\Omega$, and (b) attenuate the microwaves that are not reflected back into the cavity by 90 dB or more.

Results

The electromagnetic boundary-value problem summarized in Fig. 2 can lead to an estimate for those combinations of plasma thickness $d$ and electron number density $n_e$ which will provide sufficiently low microwave losses to permit the cavity $Q$ required. In Fig. 2, we show the *total* electromagnetic fields $E_1, H_1$ (incident + reflected) in the form of a uniform linearly polarized plane wave at the interface between the low-density positively-charged plasma region inside $r_i$ and the supercritical plasma region of thickness $d$, between $r_i$ and $r_o$. In the supercritical plasma, the electromagnetic energy represented by inward- and outward-traveling waves (summarized here as $E_2$ and $H_2$) is exponentially attenuated because the permittivity is negative. Outside the supercritical-plasma region beyond $r = r_o$, an outward-traveling electromagnetic wave represented by $E_3$ and $H_3$ can once again propagate normally. We make the simplifying assumption that the permittivity for $r<r_i$ and $r>r_o$ is $\varepsilon_0$ (free space), while the permittivity in the supercritical plasma region of thickness $d$ is $\varepsilon_0 \varepsilon_r$, where $\varepsilon_r$ is a negative value given by Eqn. (8). Another simplifying assumption (justified below) is that this region is a collisionless plasma ($\nu_{en} = 0$). With these assumptions, the boundary-value problem can be solved for the effective plasma surface resistance $R_P$ at the reflecting interface:

$$R_P = \frac{E_1}{H_1} = \eta_0 \frac{\cosh\left(d\frac{\omega}{c}\sqrt{|\varepsilon_r|}\right) + \frac{i}{\sqrt{|\varepsilon_r|}}\sinh\left(d\frac{\omega}{c}\sqrt{|\varepsilon_r|}\right)}{\cosh\left(d\frac{\omega}{c}\sqrt{|\varepsilon_r|}\right) - i\sqrt{|\varepsilon_r|}\sinh\left(d\frac{\omega}{c}\sqrt{|\varepsilon_r|}\right)} \qquad (9)$$

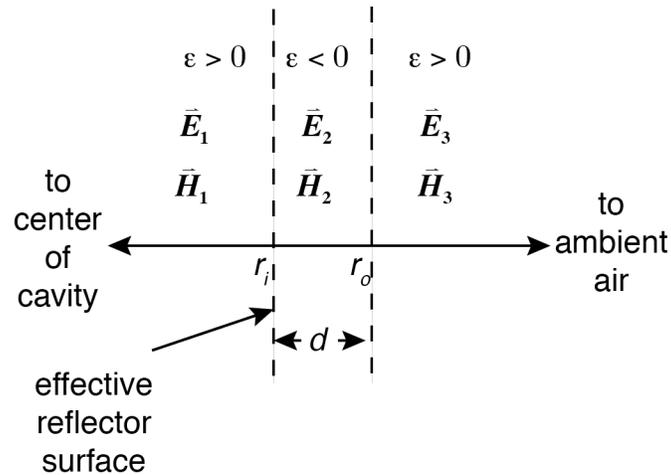

Fig. 2. Electromagnetic boundary-value problem for the supercritical plasma layer between $r_i$ and $r_o$.

The absolute value of $\varepsilon_r$ is related to the electron density $n_e$ in the dense-plasma layer according to equations (7) and (8), so the plasma's effective surface resistance at a given microwave frequency $\omega$ depends on both the thickness $d$ of the layer and its electron number density $n_e$.

Fig. 3 shows a plot of the 1-GHz cavity $Q$ that results from the plasma surface resistivity found with equation (9) as a function of $n_e$ with the plasma thickness $d$ as a parameter ranging from $d = 1$ cm to 10 cm in steps of 1 cm. Using a minimum $Q$ of $10^{10}$ as the criterion for producing a structure that could store energy long enough to account for the observed lifetime of a typical ball-lightning object, Fig. 3 shows that if $n_e = 3.5 \times 10^{11}$ cm$^{-3}$, a plasma thickness of 10 cm is necessary to provide the required reflectivity. If a higher density of $2 \times 10^{13}$ cm$^{-3}$ is present in the layer, a thickness of only 1 cm will suffice. Undoubtedly in the actual case, $n_e$ is a continuous function of radius and the boundary between the cavity and the reflecting plasma layer is not so clearly defined.

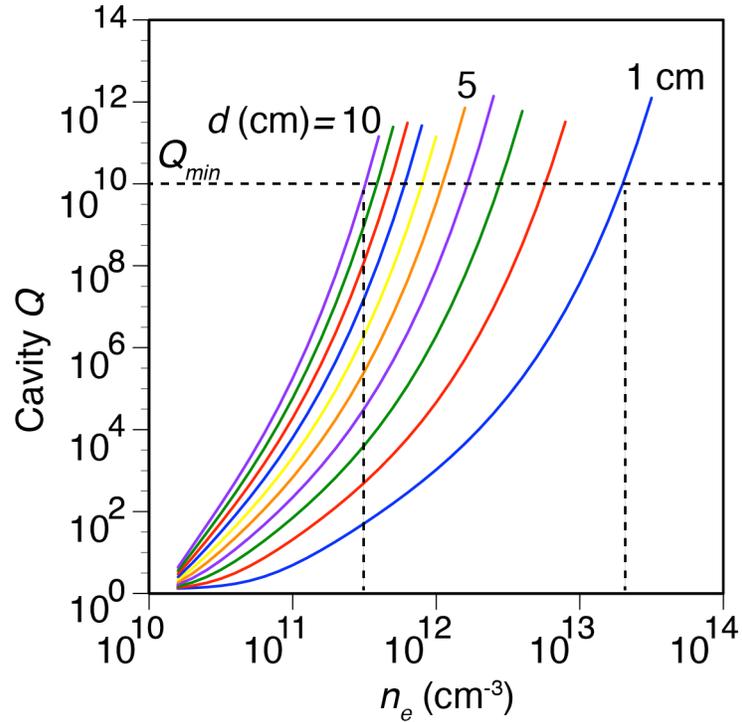

Fig. 3. Cavity $Q$ for a 1-GHz cavity as a function of electron number density $n_e$ in supercritically-dense collisionless plasma for various plasma thicknesses $d$. Vertical dashed lines indicate range of $n_e$ required for $d$ in the range 1-10 cm.

These details do not affect the main conclusion, which is that the requisite $Q$ is achievable with reasonable values of $d$ and $n_e$, as long as the plasma is collisionless or nearly so over part of its extent. A truly collisionless plasma requires that the concentration of neutral particles be much less than the electron concentration. This is not the case in most atmospheric-pressure plasmas, in which the electron-neutral collision frequency $v_{en}$ is large and would invalidate the boundary-value solution above. However, nearly all neutral atoms that stray near the outer radius $r_o$ of the object are likely to become ionized before they penetrate far into the structure. The charge separation between the inner cavity and the outer dense plasma provides a strong constant electric field directed radially outward, independent of the oscillatory fields inside the cavity. Estimates based on the charge profile of Fig. 4 (b) of Wu's paper [5] indicate that this static field may have a peak value of $10^7$ V m$^{-1}$ or more. Such a potential barrier keeps all but the most energetic positive ions out of the inner region of the dense plasma. This can justify the assumption made above that most of the supercritical plasma region can be treated as collisionless. Whether energy losses occur due to electron-neutral collisions within the supercritical plasma or due to the residual microwave field penetrating to the outside of the supercritical plasma region, the observed fact that ball-lightning objects can last for times on the order of seconds indicates that most energy-dissipation mechanisms in the supercritical-plasma region must be largely suppressed for the objects to last as long as they do, if the relativistic-microwave theory is correct.

   Normally the ball-lightning object will maintain a more or less constant radius and power emission as long as the plasma density in the supercritical plasma remains above the critical density. When that is no longer the case, the shielding provided by the supercritical plasma fails, allowing the remaining energy to radiate away. Depending on the amount of energy remaining when this happens, either the visible glow will simply disappear as the remaining energy radiates away invisibly as microwaves, or if the remaining energy is large enough, the final collapse will be accompanied by a burst of intensely ionized air as the energy dissipates in milliseconds. This behavior agrees with the two main termination modes observed in ball lightning: silent disappearance or explosion [1].

We also note that if the dominant outward pressure on the object is radiation pressure as Wu states, which varies as $r^{-3}$ for a constant total energy content, this pressure will be in a *stable* equilibrium with regard to inward pressures due to the atmosphere (constant) and to the surface tension (the latter varying as $r^{-1}$) associated with the large double-layer charge of the supercritical plasma [12]. It is worth noting in this connection that some eyewitnesses of ball lightning have likened its motion to that of a soap bubble [13, 14].

We have extended the relativistic-microwave theory of Wu to show what conditions must be satisfied by the plasma layer surrounding the proposed resonant microwave cavity in order to account for the observed lifetime of typical ball-lightning objects. We have also shown that the radial equilibrium is stable, given certain assumptions. Assuming Wu's theory to be correct, it implies very unusual characteristics for the plasma in ball lightning: an effective microwave surface resistivity on the order of $10^{-9}$ ohms per square leading to a microwave cavity $Q$ on the order of $10^{10}$. Such plasma properties, if they exist, have important scientific and technological implications. Because the microwave or electron-beam power needed to create the initial conditions assumed by this theory are currently inaccessible to convenient laboratory investigation, the entire theory cannot yet be tested. But some accessible laboratory phenomena (e. g. in fusion experiments) produce transient dense ($n_e \sim 10^{13}$ cm$^{-3}$) and nearly collisionless electron plasmas which can be investigated for their microwave reflection and transmission properties. If effective surface resistivity in the microwave region is measured which implies reflectivity better than the best-known conventional conductors, these results would confirm that the existence of such high-$Q$ naturally occurring microwave resonators as those required for this theory are at least plausible.

Methods

The data in Fig. 3 were produced with a MATLAB script which computes the cavity $Q$ using equations (6), (7), (8) and (9), from the microwave frequency $\omega$, the plasma depth $d$, and the electron concentration $n_e$.